\documentclass{moriond}
\pdfoutput=1

\usepackage{amsmath,amsfonts,amssymb}
\usepackage{multirow}
\usepackage{longtable}
\usepackage{booktabs}
\usepackage{array}

\newcommand{\C}[1]{{\cal C}_{#1}}

\def\be{\begin{equation}}
\def\ee{\end{equation}}
\def\bea{\begin{eqnarray}}
\def\eea{\end{eqnarray}}

\newcommand{\Photo}{}

\begin{document}
\vspace*{4cm}
\title{\boldmath THE $b\to s\ell^+\ell^-$ ANOMALIES\\ AND\\ THEIR IMPLICATIONS FOR NEW PHYSICS}

\author{S. DESCOTES-GENON}
\address{Laboratoire de Physique Th\'eorique, CNRS/Univ. Paris-Sud 11 (UMR 8627)\\ 91405 Orsay Cedex, France}
\author{L. HOFER \footnote{talk given by L.~Hofer at the 51st Rencontres de Moriond EW 2016}}
\address{Department de F\'isica Qu\`antica i Astrof\'isica (FQA) \\
Institut de Ci\`encies del Cosmos (ICCUB) \\
Universitat de Barcelona (UB), Mart\'i Franqu\`es 1 \\
08028 Barcelona, Spain}
\author{J. MATIAS}
\address{Universitat Aut\`onoma de Barcelona\\ 08193 Bellaterra, Spain}
\author{J. VIRTO}
\address{Theoretische Physik 1, Naturwissenschaftlich-Technische Fakult\"at\\
Universit\"at Siegen, 57068 Siegen, Germany}

\maketitle\abstracts{
Recently, the LHC has found several anomalies in exclusive semileptonic $b\to s\ell^+\ell^-$ decays. In this
proceeding, we summarize the most important results of our global analysis of the relevant decay modes. After a discussion of the hadronic uncertainties entering the theoretical predictions, we present an interpretation of the data in terms of generic new physics scenarios. To this end, we have performed model-independent fits of the corresponding Wilson coefficients to the data and have found that in certain scenarios the best fit point
is prefererred over the Standard Model by a global significance of more than 4$\sigma$. Based on the results, the discrimination between high-scale new physics and low-energy QCD effects as well as the possibility of lepton-flavour universality violation are discussed.
}

\section{Introduction}

The flavour-changing neutral current (FCNC) transition $b\to s\ell^+\ell^-$ can be probed through various decay channels, currently studied in detail at the LHC in the LHCb, CMS and ATLAS experiments, as well as at Belle. 
Recent experimental results have shown interesting deviations from the SM: The LHCb analysis~\cite{Aaij:2015oid} of the 3\,fb$^{-1}$ data on $B\to K^*\mu^+\mu^-$ in particular confirms a $\sim 3\sigma$ anomaly in two large $K^*$-recoil bins of the angular observable $P_5^\prime$~\cite{DescotesGenon:2012zf,Descotes-Genon:2013vna} that was already present in 
the 1\,fb$^{-1}$ results presented in 2013~\cite{Aaij:2013qta}. The observable 
$R_K=Br(B\to K\mu^+\mu^-)/Br(B\to Ke^+e^-)$ was measured by LHCb~\cite{Aaij:2014ora} in the dilepton mass range from 1 to 6 GeV$^2$ as $0.745^{+0.090}_{-0.074}\pm 0.036$, corresponding to a $2.6\sigma$ tension with its SM value predicted to be equal to 1 (to a very good accuracy). Finally, also the LHCb results~\cite{Aaij:2013aln} on the branching ratio of $B_s\to\phi\mu^+\mu^-$ exhibit deviations at the $\sim 3\sigma$ level in two large-recoil bins.

\begin{figure}
 \begin{minipage}{0.25\linewidth}
\begin{center}
\includegraphics[width=0.9\textwidth]{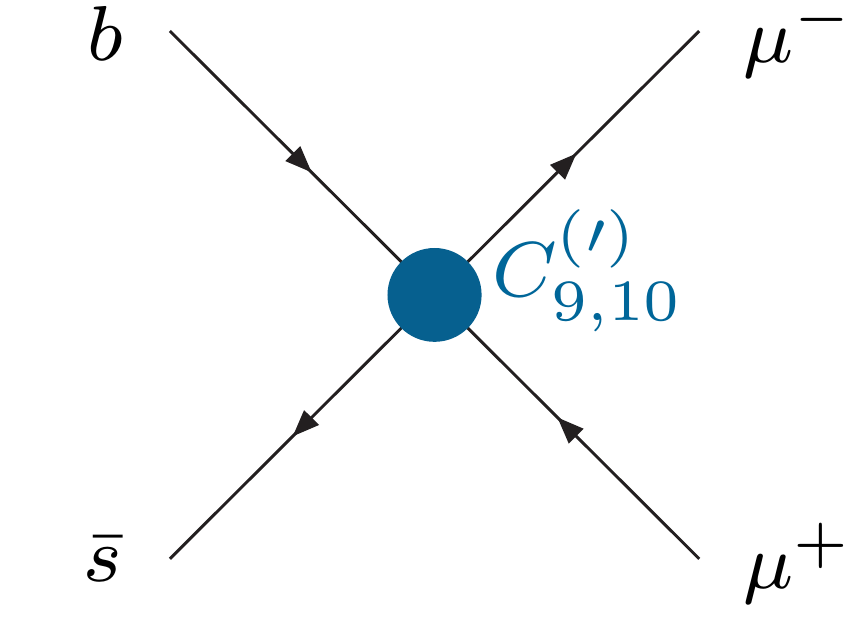}  
\end{center}
\end{minipage}
\begin{minipage}{0.25\linewidth}
\begin{center}
\includegraphics[width=0.9\textwidth]{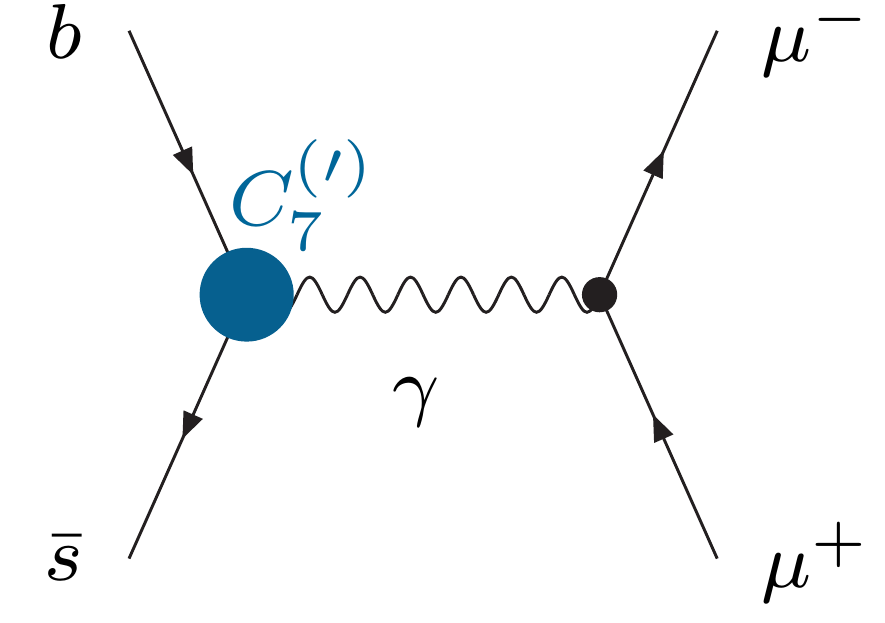}
\end{center}
\end{minipage}
\begin{minipage}{0.48\linewidth}
\begin{center}
 \begin{tabular}{c|c|c|c}
 processes &  $\C7^{(\prime)}$ &  $\C9^{(\prime)}$ &  $\C{10}^{(\prime)}$  \\\hline\hline
$B\to X_s\gamma$, $B\to K^*\gamma$ & $\checkmark$ & &  \\ \hline
$B\to X_s\mu^+\mu^-$ & $\checkmark$ & $\checkmark$ & $\checkmark$ \\ \hline
$B_s\to\mu^+\mu^-$ & & & $\checkmark$  \\ \hline
$B_{(s)}\to (K^{(*)},\phi)\mu^+\mu^-$ & $\checkmark$ & $\checkmark$ & $\checkmark$  \\ \hline
\end{tabular}
\end{center}
\end{minipage}
\caption{Effective couplings $\C{7,9,10}^{(\prime)}$ contributing to $b\to s\ell^+\ell^-$ transitions and 
sensitivity of the various radiative and (semi-)leptonic $B_{(s)}$ decays to them.
\label{fig:EffCoups}}
\end{figure}

The appearance of several tensions in different $b\to s\ell^+\ell^-$ channels is quite intriguing 
because all these observables are sensitive to the same effective couplings $\C{7,9,10}^{(\prime)}$
illustrated in Fig.~\ref{fig:EffCoups} and induced by the operators
\begin{eqnarray}
  &&\mathcal{O}_9^{(\prime)}=\frac{\alpha}{4\pi}[\bar{s}\gamma^\mu P_{L(R)}b]
   [\bar{\mu}\gamma_\mu\mu],\hspace{1cm}
  \mathcal{O}_{10}^{(\prime)}=\frac{\alpha}{4\pi}[\bar{s}\gamma^\mu P_{L(R)}b]
   [\bar{\mu}\gamma_\mu\gamma_5\mu],\nonumber\\
  &&\mathcal{O}_7^{(\prime)}=\frac{\alpha}{4\pi}m_b[\bar{s}\sigma_{\mu\nu}P_{R(L)}b]F^{\mu\nu},
\end{eqnarray}
where $P_{L,R}=(1 \mp \gamma_5)/2$ and $m_b$ denotes the $b$ quark mass. 
It is thus natural to ask whether a new physics contribution to these couplings could
simultaneously account for the various tensions in the data. Beyond the SM, contributions 
to $\C{9,10}^{(\prime)}$ are for instance generated at tree level in scenarios with $Z^\prime$ 
bosons or lepto--quarks. Note that additional scalar or pseudoscalar couplings $\C{S,S',P,P'}$ cannot address the above-mentioned anomalies since their contributions are suppressed by small lepton masses. Therefore
we will not discuss this possibility in the following.

The parameter space spanned by the couplings $\C{7,9,10}^{(\prime)}$ is probed through various observables in radiative and (semi-)leptonic $B_{(s)}$ decays, each of them sensitive to a different subset of coefficients (see Fig.~\ref{fig:EffCoups}). A complete investigation of potential new physics effects thus requires a combined study of these observables including correlations among them. The first analysis in this spirit, performed in Ref.~\cite{Descotes-Genon:2013wba} with the data of 2013, pointed to a large negative contribution to the Wilson coefficient $\C9$. This general picture was confirmed later on by other groups, using different/additional observables, different theoretical input for the form factors etc. (e.g. Refs.~\cite{Altmannshofer:2013foa,Beaujean:2013soa}). In this proceeding, we report the most important
results of our analysis in Ref.~\cite{GlobalFit} which can be compared to other recent global
analyses~\cite{Altmannshofer:2014rta,Altmannshofer:2015sma,Hurth:2016fbr} and which improves the original study in Ref.~\cite{Descotes-Genon:2013wba} in many aspects: It includes the latest experimental results of all relevant decays (using the LHCb data for the exclusive), uses refined techniques to estimate uncertainties originating from power corrections to the hadronic form factors and from non-perturbative charm loops, and consistently takes into account experimental and theoretical correlations.  

Before presenting the results from our fits in Sec.~\ref{sec:GloFi}, with a special emphasis on the possibility
of discriminating between high-scale new physics and low-energy QCD effects as well as on the possibility of lepton-flavour universality violation, we discuss in Sec.~\ref{sec:HadUn} the 
hadronic uncertainties entering the theoretical predictions of the relevant observables. 
Our conclusions are given in Sec.~\ref{sec:Conclu}.

\section{Hadronic uncertainties}
\label{sec:HadUn}

\begin{figure}
  \includegraphics[width=\linewidth]{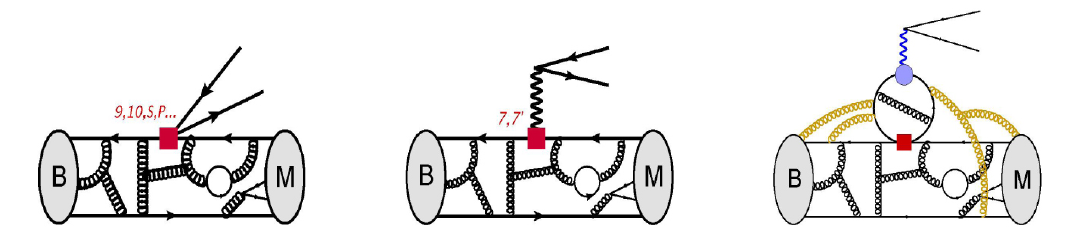}
\caption{Illustration of factorizable (first two diagrams) and non-factorizable (third diagram) 
QCD corrections to exclusive $B\to M\ell^+\ell^-$ matrix elements. \label{fig:QCDcorr}}
\end{figure}
Predictions for exclusive semileptonic $B$ decays are plagued by QCD effects of perturbative and non-perturbative nature. At leading order (LO) in the effective theory, predictions involve tree-level diagrams with insertions of the 
operators $\mathcal{O}_{7,9,10}$ (generated at one loop in the SM), as well as one-loop diagrams
with an insertion of the charged-current operator $\mathcal{O}_2=[\bar{s}\gamma^\mu P_Lc][\bar{c}\gamma_\mu P_Lb]$
(generated at tree level in the SM). In contributions of the first type,
the leptonic and the hadronic currents factorize, and QCD corrections are constrained to the hadronic $B\to M$ current (first two diagrams in Fig.~\ref{fig:QCDcorr}). This class of {\it factorizable QCD corrections} thus forms part of the hadronic form factors parametrizing the $B\to M$ transition.
Contributions of the second type, on the other hand, receive {\it non-factorizable QCD corrections} 
(third diagram in Fig.~\ref{fig:QCDcorr}) that cannot be absorbed into form factors. In the following
we discuss the uncertainties stemming from the two types of corrections and their implementation in our analysis.

\subsection{Form factor uncertainties}
The form factors are available from lattice as well as from light-cone sum rule (LCSR) calculations, with the former
being suited for the region of high $q^2>15$\,GeV$^2$ and the latter for the region of low $q^2<8$\,GeV$^2$. Since the form factors introduce a dominant source of uncertainties into the theory predictions, it is desirable to reduce the sensitivity to them as much as possible. For $B\to V\ell^+\ell^-$ decays, with $V$ being a vector meson, 
this can be achieved in the low-$q^2$ region by exploiting large-recoil symmetries of QCD. 
At LO in $\alpha_s$ and $\Lambda/m_b$, these symmetries enforce certain relations among the seven hadronic 
form factors $V$, $A_1$, $A_2$, $A_0$, $T_1$, $T_2$, $T_3$, like e.g.
\begin{equation}
\frac{m_B(m_B+m_{K^*})A_1-2E(m_B-m_{K^*})
A_2}{m_B^2T_2-2Em_BT_3}\,=\,1+\mathcal{O}(\alpha_s,\Lambda/m_b),
\label{eq:ratio}
\end{equation}
where $m_B$ denotes the mass of the $B$ meson, and $m_{K^*}$ and $E$ the mass and the energy of 
the $K^*$ meson. From the experimentally measured coefficients of the differential angular distribution of $B\to V\ell^+\ell^-$, one
can construct observables that involve ratios like the one in eq.~(\ref{eq:ratio}). The resulting observables
$P_i^{(\prime)}$ then only exhibit a mild form factor dependence, suppressed by powers of $\alpha_s$ 
and $\Lambda/m_b$.

For the cancellation of the form factor uncertainties in ratios like the one in eq.~(\ref{eq:ratio}), it is crucial
to have control of the correlations among the errors of the different form factors. These correlations can be
taken into account via two orthogonal approaches: Either they can be assessed directly from the LCSR calculation
(Ref.~\cite{Straub:2015ica} provides LCSR form factors with correlation matrices), or they can be implemented resorting to the 
large-recoil symmetry relations. Whereas the former method is limited to the particular set of
LCSR form factors from Ref.~\cite{Straub:2015ica} and hence sensitive to details of the corresponding calculation, the latter method determines the correlations in a model-independent way from first principles and can thus 
also be applied to different sets of form factors like the ones from Ref.~\cite{Khodjamirian:2010vf},
where no correlations were provided. As a drawback, correlations are obtained from large-recoil symmetries only up to $\Lambda/m_b$ corrections which have to be estimated. For the estimate of these {\it factorizable power corrections}, we follow the strategy that was developed in Ref.~\cite{Descotes-Genon:2014uoa} based on and further refining a method first proposed in Ref.~\cite{Jager:2012uw}.
We assume a generic size of $10\%$ factorizable power corrections to the form factors, which is consistent
with the results that are obtained from a fit to the particular LCSR form factors from Refs.~\cite{Khodjamirian:2010vf,Straub:2015ica}.

\subsection{Uncertainties from $c\bar{c}$ loops}
Long-distance charm-loop effects (third diagram in Fig.~\ref{fig:QCDcorr}) can mimic the effect of
an effective coupling $\C9^{c\bar{c}}$ and have been suggested as a solution of
the anomaly in $B\to K^*\mu^+\mu^-$ \cite{Lyon:2014hpa,Ciuchini:2015qxb}. Due to the non-local structure of these corrections, their 
contribution is expected to have a non-constant $q^2$-dependence, where $q^2$ is the squared invariant masses
of the lepton pair. Together with the perturbative SM contribution $\C{9\;\rm SM\,pert}^{\rm eff}$ 
and a potential constant new physics coupling $\C9^{\rm SM}$, it can be cast into an effective Wilson coefficient
\begin{equation}
 \C9^{\rm eff\;i}(q^2)\,=\,\C{9\;\rm SM\,pert.}^{\rm eff}(q^2)\,+\,\C9^{\rm NP}\,+\,
  \C9^{c\bar{c}\;i}(q^2),
 \label{eq:EffCharm}
\end{equation}
with a different $\C9^{c\bar{c}\;i}$ and hence also a different $\C9^{\rm eff\;i}$ for the three transversity amplitudes $i=0,\|,\perp$. Currently, only a partial calculation~\cite{Khodjamirian:2010vf} exists, yielding values 
$\C{9\;\rm KMPW}^{c\bar{c}\;i}$ that tend to enhance the anomalies. In our analysis, we assume that this 
partial result is representative for the order of magnitude of the total charm-loop contribution and
we assign an error to unknown charm-loop effects varying
\begin{equation}
  \C9^{c\bar{c}\;i}(q^2)\;=\;s_i\;\C{9\;\rm KMPW}^{c\bar{c}\;i}(q^2),\hspace{0.5cm}
  \textrm{for }-1\le s_i\le 1.
\end{equation}

\section{Results of the global fit}
\label{sec:GloFi}

Our reference fits are obtained using the following experimental input: branching ratios and angular observables
of the decays $B\to K^*\mu^+\mu^-$ and $B_s\to\phi\mu^+\mu^-$, branching ratios of the charged and neutral modes
$B\to K\mu^+\mu^-$, the branching ratios of $B\to X_s\mu^+\mu^-$, $B_s\to\mu^+\mu^-$ and $B\to X_s\gamma$,
as well as the isospin asymmetry $A_I$ and the time-dependent CP asymmetry $S_{K^*\gamma}$ of $B\to K^*\gamma$.
For the theoretical predictions, we use lattice form factors from Refs.~\cite{Horgan:2013hoa,Bouchard:2013pna} 
in the low-recoil region, and LCSR form factors from Ref.~\cite{Khodjamirian:2010vf} 
(except for $B_s\to\phi$ where Ref.~\cite{Straub:2015ica} is used), with correlations 
assessed from the large-recoil symmetries.
  
\begin{table}
\begin{center}
\begin{tabular}{@{}crccc@{}}
\toprule[1.6pt] 
Coefficient & Best fit & 1$\sigma$ & 3$\sigma$ & Pull$_{\rm SM}$ \\ 
 \midrule 
 $\C7^{\rm NP}$ & $ -0.02 $ & $ [-0.04,-0.00] $ & $ [-0.07,0.03] $ &  1.2 \\[2mm] 
 \boldmath$\C9^{\rm NP}$ & \boldmath$ -1.09 $ & $ [-1.29,-0.87] $ & $ [-1.67,-0.39] $ &  \bf 4.5 \\[2mm] 
 $\C{10}^{\rm NP}$ & $ 0.56 $ & $ [0.32,0.81] $ & $ [-0.12,1.36] $ &  2.5 \\[2mm] 
 $\C{7'}^{\rm NP}$ & $ 0.02 $ & $ [-0.01,0.04] $ & $ [-0.06,0.09] $ &  0.6 \\[2mm] 
 $\C{9'}^{\rm NP}$ & $ 0.46 $ & $ [0.18,0.74] $ & $ [-0.36,1.31] $ &  1.7 \\[2mm] 
 $\C{10'}^{\rm NP}$ & $ -0.25 $ & $ [-0.44,-0.06] $ & $ [-0.82,0.31] $ &  1.3 \\[2mm] 
 $\C9^{\rm NP}=\C{10}^{\rm NP}$ & $ -0.22 $ & $ [-0.40,-0.02] $ & $ [-0.74,0.50] $ &  1.1 \\[2mm] 
 \boldmath$\C9^{\rm NP}=-\C{10}^{\rm NP}$ & \boldmath$ -0.68 $ & $ [-0.85,-0.50] $ & $ [-1.22,-0.18] $ &  \bf 4.2 \\[2mm] 
 \boldmath$\C9^{\rm NP}=-\C{9'}^{\rm NP}$ & \boldmath$ -1.06 $ & $ [-1.25,-0.85] $ & $ [-1.60,-0.40] $ &  \bf 4.8 \\[2mm] 
\bottomrule[1.6pt] 
\end{tabular}
\end{center}
\caption{Results of various one-parameter fits for the Wilson coefficients $\{\C{i}\}$.
\label{tab:fitres}}
\end{table}

Starting from a model hypothesis with $n$ free parameters for the Wilson coefficients $\{\C{i}^{\rm NP}\}$, we then
perform a frequentist $\Delta\chi^2$-fit, including experimental and theoretical correlation matrices.
In Tab.~\ref{tab:fitres} we show our results for various one-parameter scenarios. In the last column 
we give the SM-pull for each scenario, i.e. we quantify by how many sigmas the best fit point is preferred
over the SM point $\{\C{i}^{\rm NP}\}=0$ in the given scenario. A scenario with a large SM-pull thus allows for
a big improvement over the SM and a better description of the data. From the results in 
Tab.~\ref{tab:fitres} we infer that a large negative $\C9^{\rm NP}$ is required to explain the data. In a scenario where only
this coefficient is generated a fairly good goodness-of-fit is yielded for $\C9^{\rm NP}\sim-1.1$. A decomposition into the
different exclusive decay channels, as well as into low- and large-recoil regions, shows that each 
of these individual contributions points to the same solution, i.e. a negative $\C9^{\rm NP}$, albeit with 
varying significance. We refer the reader to Ref.~\cite{GlobalFit} for further results, e.g. for fits in various
2-parameter scenarios as well as for the full 6-parameter fit of $\C{7,9,10}^{(\prime)\rm NP}$ 
resulting in a SM-pull of $3.6\sigma$. 

\begin{figure}
\begin{center}
\begin{minipage}{0.55\linewidth}
  \includegraphics[width=1.1\linewidth]{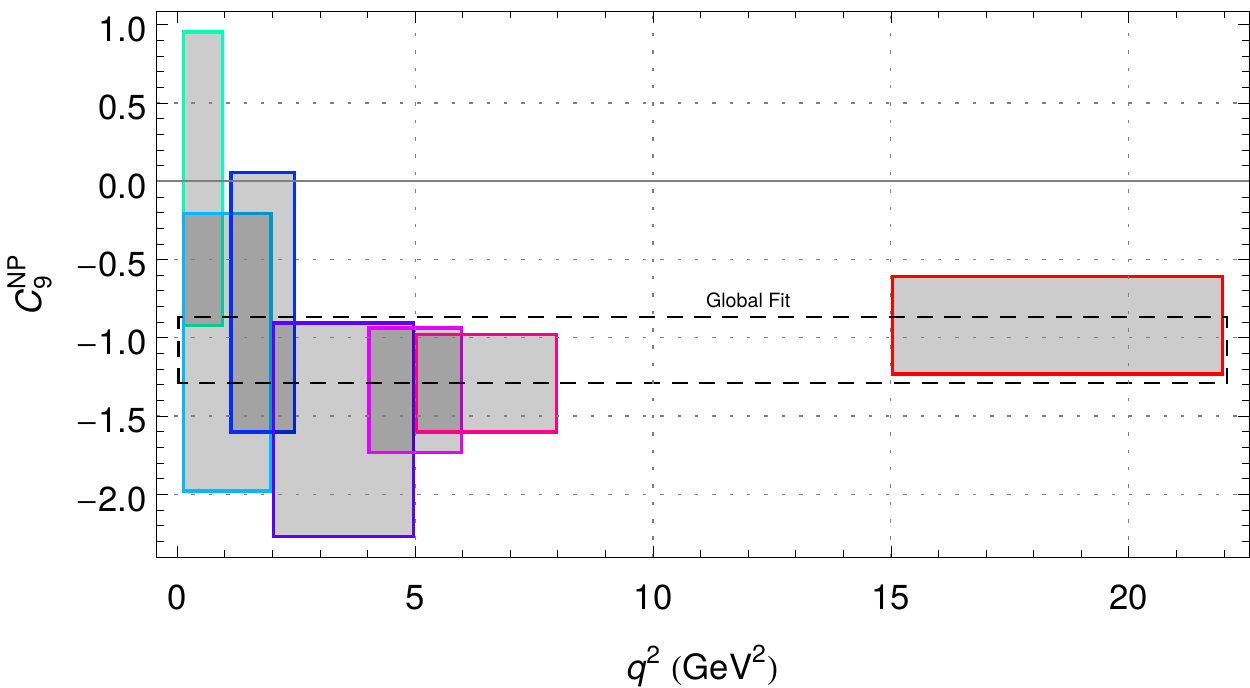}
\end{minipage}\hspace{0.08\linewidth}
\begin{minipage}{0.35\linewidth}
 \includegraphics[width=0.95\linewidth]{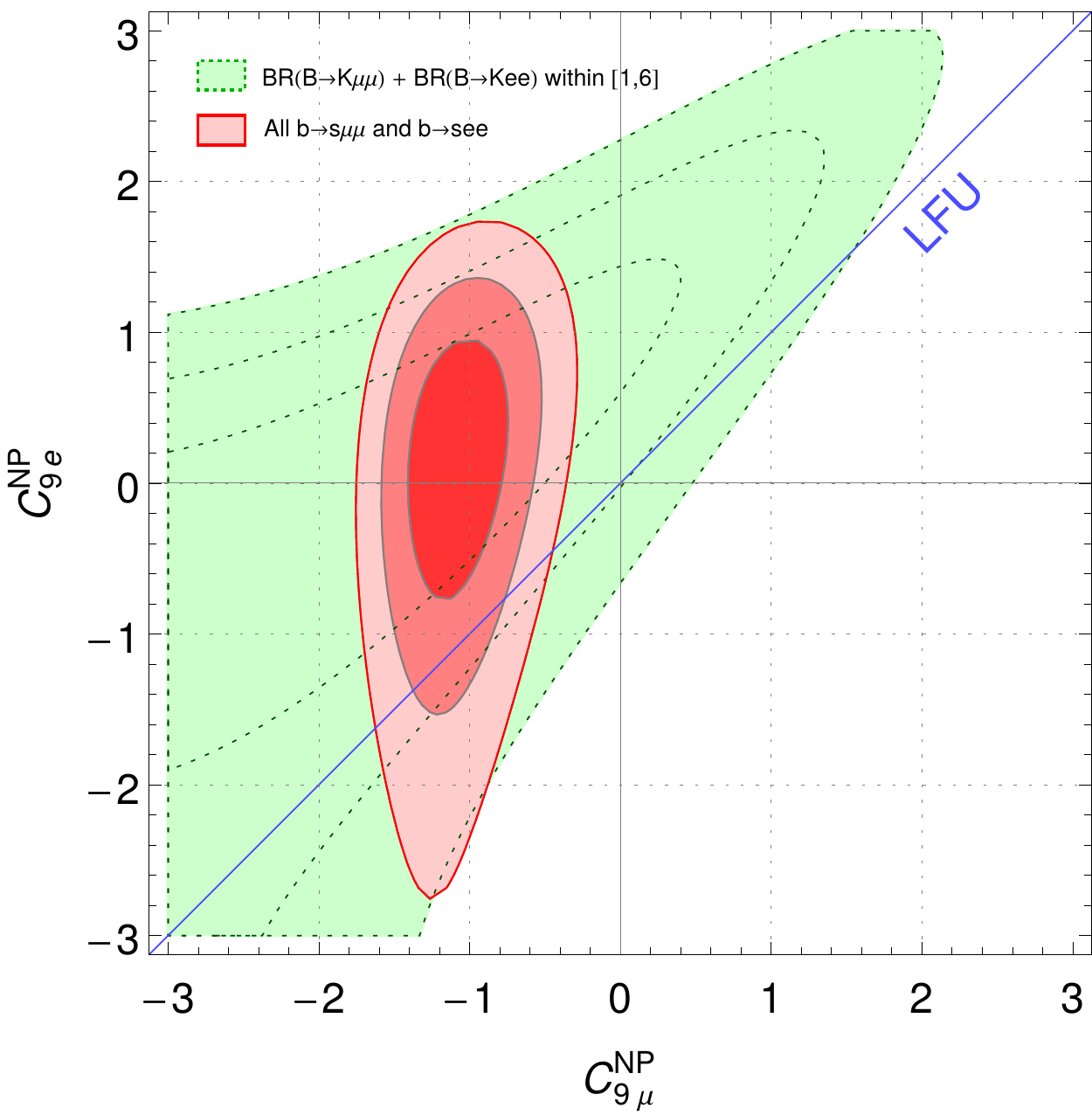}
\end{minipage} 
\end{center}
\caption{Left: Bin-by-bin fit of the one-parameter scenario with a single coefficient $\C9^{\rm NP}$.
         Right: Fit with independent coefficients $\C{9\;\mu}^{\rm NP}$ and $\C{9\;e}^{\rm NP}$.
\label{fig:BinFit}}
\end{figure}

\subsection{New physics vs. non-perturbative charm-contribution}
According to Eq.~(\ref{eq:EffCharm}), a potential new physics contribution $\C9^{\rm NP}$ enters amplitudes  
always together with a charm-loop contribution $\C9^{c\bar{c}\;i}(q^2)$, spoiling an unambiguous
interpretation of the fit result from the previous section in terms of new physics. However, whereas
$\C9^{\rm NP}$ does not depend on the squared invariant mass $q^2$ of the lepton pair, $\C9^{c\bar{c}\;i}(q^2)$
is expected to exhibit a non-trivial $q^2$-dependence. Following Ref.~\cite{Altmannshofer:2015sma},
we show in Fig.~\ref{fig:BinFit} on the left a bin-by-bin fit for the one-parameter scenario 
with a single coefficient $\C9^{\rm NP}$. The results obtained in the individual bins are consistent with each 
other, allowing thus for a solution $\C9^{\rm NP}$ that is constant in the whole $q^2$ region, as required for an interpretation in terms of new physics, though the situation is not conclusive due to the large uncertainties in the single bins.

An alternative strategy to address this question has been followed recently in Ref.~\cite{Ciuchini:2015qxb} where a direct fit of the $q^2$-dependent 
charm contribution $\C9^{c\bar{c}\;i}(q^2)$ to the data on $B\to K^*\mu^+\mu^-$ (at low $q^2$) has been performed under the hypothesis of the absence of new physics.
The fact that only one out of the 12 parameters encoding a non-constant $q^2$-dependence in $\C9^{c\bar{c}\;i}(q^2)$ differs from zero by $\gtrsim 2\sigma$
makes the result obtained in Ref.~\cite{Ciuchini:2015qxb} compatible \footnote{The probability for a fluctuation by $\gtrsim 2\sigma$ in at least one out of 12 parameters is given by 
$1-0.954^{12}\approx 43\%$, well within $1\sigma$.} 
with a $q^2$-independent new physics solution $\C9^{\rm NP}$, in agreement with our findings from Fig.~\ref{fig:BinFit}. 
Note further that the results in Ref.~\cite{Ciuchini:2015qxb} do not allow to draw any conclusions on whether a $q^2$-dependent solution of the anomalies via  $\C9^{c\bar{c}\;i}(q^2)$
is preferred compared to a solution via a constant $\C9^{\rm NP}$ since this
would require a comparison of the goodness of the fit taking into account the different number of free parameters of
the two parametrizations. Moreover, we like to stress that the results for the observables presented in Ref.~\cite{Ciuchini:2015qxb} should not be interpreted as SM predictions, 
as they are based on a fit to the experimental data.

\subsection{Lepton-flavour universality violation}
Since the measurement of $R_K$ suggests the violation of lepton-flavour universality, we also
studied the situation where the muon- and the electron-components of the operators
$\C{9,10}^{(\prime)}$ receive independent new physics contributions $\C{i\;\mu}^{\rm NP}$ and
$\C{i\;e}^{\rm NP}$, respectively. The electron-couplings $\C{i\;e}^{\rm NP}$ are constrained by adding
the decays $B\to K^{(*)}e^+e^-$ to the global fit. Note that the correlated fit to 
$B\to K\mu^+\mu^-$ and $B\to Ke^+e^-$ simultaneously is equivalent to a direct inclusion of the observable $R_K$.

In Fig.~\ref{fig:BinFit} on the right we display the result for the two-parameter fit to the coefficients
$\C{9\;\mu}^{\rm NP}$ and $\C{9\;e}^{\rm NP}$. The fit prefers an electron-phobic 
scenario with new physics coupling to $\mu^+\mu^-$ but not to $e^+e^-$. Under this hypothesis,
that should be tested by measuring $R_{K^*}$ and $R_\phi$,
the SM-pull increases by $\sim 0.5\sigma$ compared to the value in Tab.~\ref{tab:fitres} for the lepton-flavour universal scenario.

\section{Conclusions}
\label{sec:Conclu}

LHCb data on $b\to s\ell^+\ell^-$ decays shows several tensions with SM predictions, in particular in the angular observable $P_5^\prime$ of $B\to K^*\mu^+\mu^-$, 
in the branching ratio of $B_s\to\phi\mu^+\mu^-$, and in the ratio
$R_K=Br(B\to K\mu^+\mu^-)/Br(B\to Ke^+e^-)$ (all of them
at the $\sim 3\sigma$ level). In global fits of the Wilson coefficients to the data,
scenarios with a large negative $\C9^{\rm NP}$ are preferred over the SM by typically more than $4\sigma$.
A bin-by-bin analysis demonstrates that the fit is compatible with a $q^2$-indepedent effect generated by high-scale new physics, though a $q^2$-dependent QCD effect cannot be excluded with the current precision. Note, however, that a QCD effect could not explain the tension in $R_K$. The latter observable further favours a lepton-flavour violating scenario with new physics coupling only to $\mu^+\mu^-$ but not to $e^+e^-$, a scenario to be probed by a measurement of the analogous ratios $R_{K^*}$ and $R_\phi$ to probe this hypothesis.

\section*{Acknowledgments}

L.H. is grateful to the organizers for the invitation to the workshop and thanks the participants
for stimulating discussions. The work of L.H.\ was supported by the grants FPA2013-46570-C2-1-P and 2014-SGR-104, and partially by the Spanish MINECO under the project MDM-2014-0369 of
ICCUB (Unidad de Excelencia ``Mar\'ia de Maeztu''). JV is funded by the DFG within research unit FOR 1873 (QFET), and acknowledges financial support from CNRS. SDG, JM and JV acknowledge financial support from FPA2014-61478-EXP.
This project has received funding from the European Union’s Horizon 2020 research and innovation programme under grant agreements No 690575, No 674896 and No. 692194.

\end{document}